\documentclass[aps,prl,superscriptaddress,showpacs,floats,twocolumn,floatfix]{revtex4-1}

\usepackage{bm}
\usepackage{amsmath,amsfonts}
\usepackage{graphicx}
\usepackage{color}
\usepackage[usenames,dvipsnames,svgnames,table]{xcolor}
\usepackage{hyperref}
\usepackage{setspace}
\usepackage{soul}
\usepackage{accents}
\usepackage{mathtools}

\makeatletter
\def\squiggly{\bgroup \markoverwith{\textcolor{red}{\lower3.5\p@\hbox{\sixly \char58}}}\ULon}
\makeatother

\newcommand{\beq}{\begin{equation}}
\newcommand{\eeq}{\end{equation}}
\newcommand{\beqa}{\begin{eqnarray}}
\newcommand{\eeqa}{\end{eqnarray}}
\newcommand{\beqan}{\begin{eqnarray*}}
\newcommand{\eenan}{\end{eqnarray*}}

\newcommand\defeq{:{\kern -0.5em}=}

\newcommand{\km}{ {\cal K}}

\newcommand{\ww}{{\bm \omega}}
\newcommand{\vv}{{\bm v}}
\newcommand{\pp}{{\bm p}}
\newcommand{\unitp}{\hat{p}}
\newcommand{\unitv}{\hat{v}}
\newcommand{\unitw}{\hat{\omega}}
\newcommand{\unitx}{\hat{x}}
\newcommand{\unity}{\hat{y}}
\newcommand{\unitz}{\hat{z}}
\newcommand{\Dort}{ D_{\text{o}} }

\newcommand{\Deff}{ D_{\mbox{\scriptsize eff}} }
\newcommand{\repere}{{\bm{\mathsf E}}}
\newcommand{\Drinf}{\langle\Delta{\bm r}\rangle_\infty}

\begin{document}

\addtolength{\topmargin}{10pt}

\title{Spiral diffusion of rotating self-propellers with stochastic perturbation}
\author{Amir Nourhani}
\email{nourhani@psu.edu}
\affiliation{Center for Nanoscale Science,The Pennsylvania State University, University Park, PA 16802, USA}
\affiliation{Department of Physics, The Pennsylvania State University, University Park, PA 16802, USA} 
\author{Stephen J. Ebbens}
\affiliation{Department of Chemical \& Biological Engineering, University of Sheffield, UK}
\author{John G. Gibbs}
\affiliation{Department of Physics and Astronomy, Northern Arizona University, Flagstaff, Arizona 86011, USA}
\affiliation{Center for Bioengineering Innovation, Northern Arizona University, Flagstaff, Arizona 86011, USA}
\author{Paul E. Lammert}
\email{lammert@psu.edu}
\affiliation{Center for Nanoscale Science,The Pennsylvania State University, University Park, PA 16802, USA}
\affiliation{Department of Physics, The Pennsylvania State University, University Park, PA 16802, USA}

%\date{\today}

\begin{abstract}
Translationally diffusive behavior arising from the combination of orientational 
diffusion and powered motion at microscopic scales is a known phenomenon, but 
the peculiarities of the evolution of expected position conditioned on initial
position and orientation have been neglected. 
A theory is given of the spiral motion of the mean trajectory depending 
upon propulsion speed, angular velocity, orientational diffusion and
rate of random chirality reversal. We demonstrate the experimental accessibility
of this effect using both tadpole-like and Janus sphere dimer rotating motors.
Sensitivity of the mean trajectory to the kinematic parameters suggest that
it may be a useful way to determine those parameters.
\end{abstract}

\maketitle

Active colloids such as microswimmers and nanomotors are a class of 
non-equilibrium systems 
which has been the subject of intense research in recent years~\cite{KimNanoscale2016,VolpeArXiv2016,ReichhardtArXiv2016}.
At the sub-micron length scale, stochastic effects significantly
perturb a self-propeller's deterministic motion, and the coupling of such
noise to a steady motion can lead to unexpected emergent phenomena 
such as motility-induced phase separation~\cite{Tailleur+Cates-08},
chiral diffusion~\cite{Nourhani+13-PRE}, and
phenomena with biological relevance~\cite{Friedrich+Julicher-08} 
which can now be modelled by artificial active colloids. 
In the absence of noise a circle swimmer, confined to a plane with a strong rotational component to 
its powered motion, travels on a fixed circle with a steady clockwise or 
counterclockwise chirality~\cite{nourhani2013p062317}.
Artificial swimmers of this sort have been fabricated in a variety of 
forms such as tadpoles~\cite{Gibbs+Zhao-10,Gibbs+11}, 
Janus sphere dimers~\cite{Ebbens+10,Valadares+10}, 
nanorods~\cite{Takagi:2013,Wang:2009p80}, 
and acoustically-activated swimmers~\cite{Ahmed+15}. 
Stochastic perturbations in the form of unbiased orientational diffusion
or random chirality-reversal resulting from flipping about the direction of 
motion have significant effects on the long term motion:
an effective translational diffusion is generated~\cite{MarchesoniSoftMatter2016,MarchesoniPRE2016,LowenPRL2013Spiral},
the infinite-time limit of the mean position conditioned on the initial position and velocity
is non-zero and chirality-dependent~\cite{Nourhani+13-PRE},
and the mean approach to the limit is a logarithmic spiral~\cite{VanTeeffelen,LowenPRL2013Spiral}.

\begin{figure}[b]
\begin{center}
\includegraphics[width=2.8in]{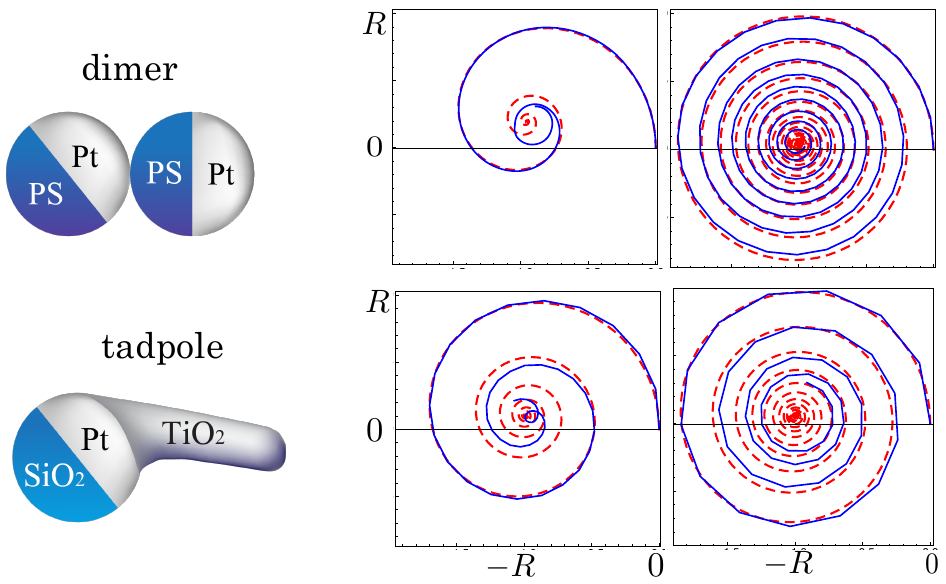}
\end{center}
\vspace{-15pt}
\caption{
\label{fig:spirals} 
(color online)
Traces of 
mean trajectories (solid blue) of synthetic ensembles constructed from video of a
single motor trajectory, along with fits to the theory (dashed red).
Two trajectories for each of the motor types, tadpoles and Janus sphere dimers
were selected for this investigation. The fit parameters 
$(\omega\, [\text{rad}/\text{s}], \Dort\, [\text{rad}^2/\text{s}])$ to trajectories
for dimers are left: (0.86, 0.176) and right: (1.07, 0.037),
and for tadpole-like swimmers are left: (4.26, 0.446) and right: (6.15, 0.321).
\label{fig:expSpiral}
}
\end{figure}

In this Letter, we  experimentally and theoretically 
demonstrate ``spiral diffusion'' as a general finite-time behavior of the conditional
mean position in circle swimmers. First, we expose the phenomenon in experimental data for 
both tadpole-like~\cite{Gibbs+11} and Janus-sphere dimer~\cite{Ebbens+10} rotary 
microswimmers (see Fig.~\ref{fig:expSpiral}), and present fits to the model.
Then, we explain the theory for spiral diffusion of circle swimmers subjected to both 
orientational diffusion and flipping (change of chirality).
The expected position of the swimmer, conditioned on its initial position, velocity direction and
chirality, evolves along a converging spiral.
The theory serves as a sensitive and accurate utility for determining kinematic 
parameters such as angular velocity and orientational diffusivity.
Supplementary Material contains the details of fabrication and experimental protocols, as well as movies of
simulated ensembles of swimmers for a variety of noise parameters.

%%%%%%%%%%%
%%%%%%%%%%%%%%

Experiments were performed on two different rotor designs, 
tadpole-like microswimmers~\cite{Gibbs+11} and Janus sphere dimers~\cite{Ebbens+10}.
The motors are denser than the aqueous solution of hydrogen peroxide,
thus they move near the substrate and effectively confined to a horizontal plane.
Even within a batch of nominally identical motors, there is usually
a significant range of kinematic parameters. 
The analyzed experimental data consisted of two 
videos for each type of swimmer.
From a single video of $N$ frames, an ensemble
of $N_{\text{traj}}$ trajectories, each of length $N-N_{\text{traj}}$ frames
is synthesized. For $1\le n \le N_{\text{traj}}$, 
the $n$-th member of the ensemble is obtained by taking
frames $n$ through $N - N_{\text{traj}} + n - 1$ of the original video and 
rotating them so that the initial velocities are always in the 
same ($\unitv_0$) direction.
Average positions of these synthesized ensembles are shown as solid blue 
spirals in Fig.~\ref{fig:spirals}.
Fits to the theory, as explained below, are shown as dashed 
red curves, and are obtained by adjusting the angular speed $\omega$,
linear speed $v = R \omega$, 
and orientational diffusion coefficient $\Dort$. With this method,
we find a much more sensitive fit to 
these kinematic parameters
than from working directly with
the orientation time series 
and mean square displacement;
a small change 
in the ratio $\Dort/\omega$ can change the shape of the spiral. 

%%%%%%%%%%%
\begin{figure}[t]
\begin{center}
\includegraphics[width=3.3in]{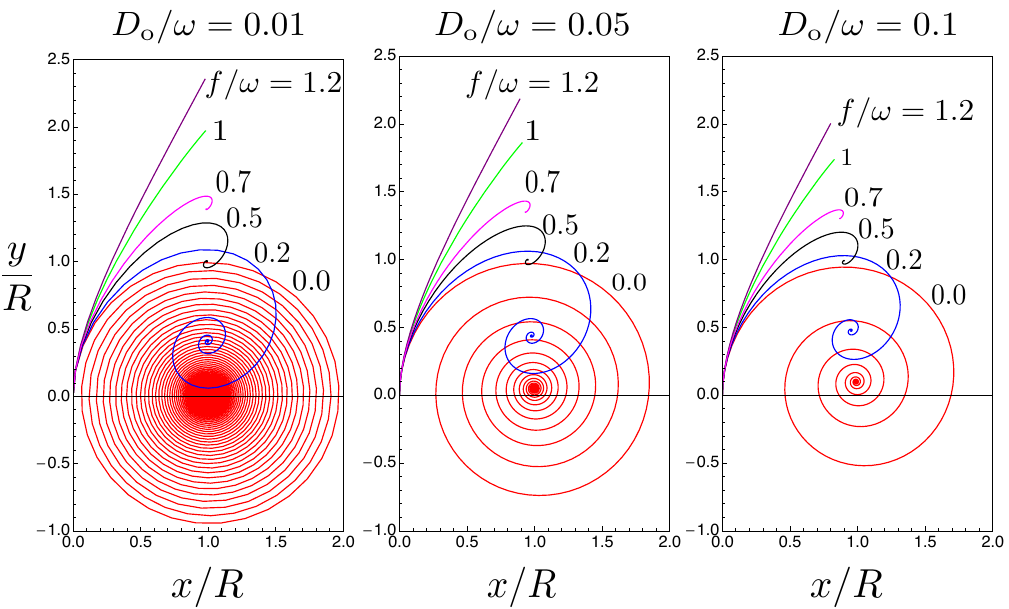}
\end{center}
\vspace{-15pt}
\caption{
\label{fig:spirals2} 
(color online)
 Traces
of the mean position $\langle {\bm r}(t)\rangle$ 
of clockwise rotary self-propellers
conditioned
on initial position at the origin and initial velocity directed along
$\hat{e}_y$, for 
dimensionless
orientational diffusivity $\Dort/\omega = 0.01,0.05$ and $0.1$
and
dimensionless
 flipping rate $f/\omega = 0, 0.2, 0.5, 0.7, 1$ and $1.2$.
According to Eqs. (\ref{eq:displacementvecRT}),
(\ref{eq:asymptotic-mean-displacement}), and (\ref{eq:G-vector}),
$\langle{\bm r}(t)\rangle$ spirals in to its asymptotic value when
$f < \omega$, but the approach is non-oscillatory for $f > \omega$.
Watch the supplementary video ``vid-fig2.m4v'' for a combination of theory and simulation.
\label{fig:asympt}
}
\end{figure}
%%%%%%%%%%%%%%
To develop the theory, we begin with the deterministic part of a circle swimmer's 
motion.
The particle
 moves with constant linear $\vv= v \unitv$ and angular $\ww= \omega \unitw$ ($\omega\geq 0$) velocities; The instantaneous orbit of motion has radius $R = v/\omega$ and the vector 
$\pp = R\, \unitp = R \unitv\times\unitw$ 
connects the center of instantaneous orbit to the self-propeller. 
Assuming the particles start from the same initial position and 
velocity, the time-dependent right-handed body frame
${\repere}(t) = [\unitp, \unitv, \unitw]^T(t)$ 
is related to the fixed laboratory frame by 
$\repere(0) = [\unitp_0, \unitv_0, \unitw_0]^T = [\unitx, \unity, \unitz]^T$ for counterclockwise rotation,
($\repere(0) = [-\unitx, \unity, -\unitz]^T$ for clockwise).
To study the dynamics of these particles we use the kinematrix theory~\cite{Nourhani+14-PRE,Nourhani+-14-Gaussian}, 
that we recently developed as an alternative to Langevin and Fokker-Planck formalisms. 
In the limit of short noise correlation and momentum relaxation times,
the self-propeller's kinematic properties such as orientational diffusion, angular 
speed and flipping rate can be packaged into 
a $3\times3$ {\em kinematrix} ${\cal K}$.
The dynamics of the body frame is governed by
$\frac{d}{dt}\langle \repere(t)\rangle = -{\cal K}\langle\repere(t)\rangle$
where $\langle \cdot\rangle$ is the ensemble average operator over all realization of noises.
This model is appropriate to nanomotors and microswimmers at low Reynolds number,
since the relaxation time due to viscous damping is very short (for a micron-sized object,
of order  1 $\mu$s), and correlation times of environmental stochastic forces are even shorter.

The self-propeller  moves near a plane in 2D 
($\unitv \perp\unitw$), undergoing orientational diffusion with diffusivity 
$D_\text{o}$ about $\unitw$ while,
simultaneously and independently, it flips about its direction of motion $\unitv$ 
with frequency $f$ and thereby reversing chirality. Although the motors in our experimental study had stable chirality, some artificial motors 
may experience  flipping~\cite{Takagi:2013}.
The kinematrix for this model is~\cite{Nourhani+14-PRE}
\begin{equation}
\km
=
 \begin{bmatrix}
   D_\text{o}\!+\!2f & \omega             &   0  \\ 
   -\omega         & D_\text{o}         &   0  \\  
   0                    & 0                       &  2f
\end{bmatrix}.
\label{eq:RTkinematrix}
\end{equation}
The stochastic motion of the 
body
frame generates an effective (long-time) translational 
diffusivity
\begin{equation}
\label{eq:DeffMinverse}
\Deff
= {v^2 \over 2}\left[\km^{-1}\right]_{22}
=
\frac{\omega R^2}{2} 
\frac{\omega(\Dort + 2 f)}{(\Dort + f)^2 + (\omega^2 - f^2)}.
\end{equation}
Passive translational diffusion with diffusivity $D_t$ contributes an independent 
diffusion, so that the net diffusion coefficient is $\Deff + D_t$.
Passive translational diffusion is not coupled to the powered motion, but
orientational diffusion is. It is for this reason that the latter
can dominate the total diffusion. 
Random flipping instantaneously creates a large qualitative change in the motion
and is alone sufficient to generate long-term diffusion.

%%%%%%%%%%%
\begin{figure*}
\begin{center}
\includegraphics[width=17.3cm]{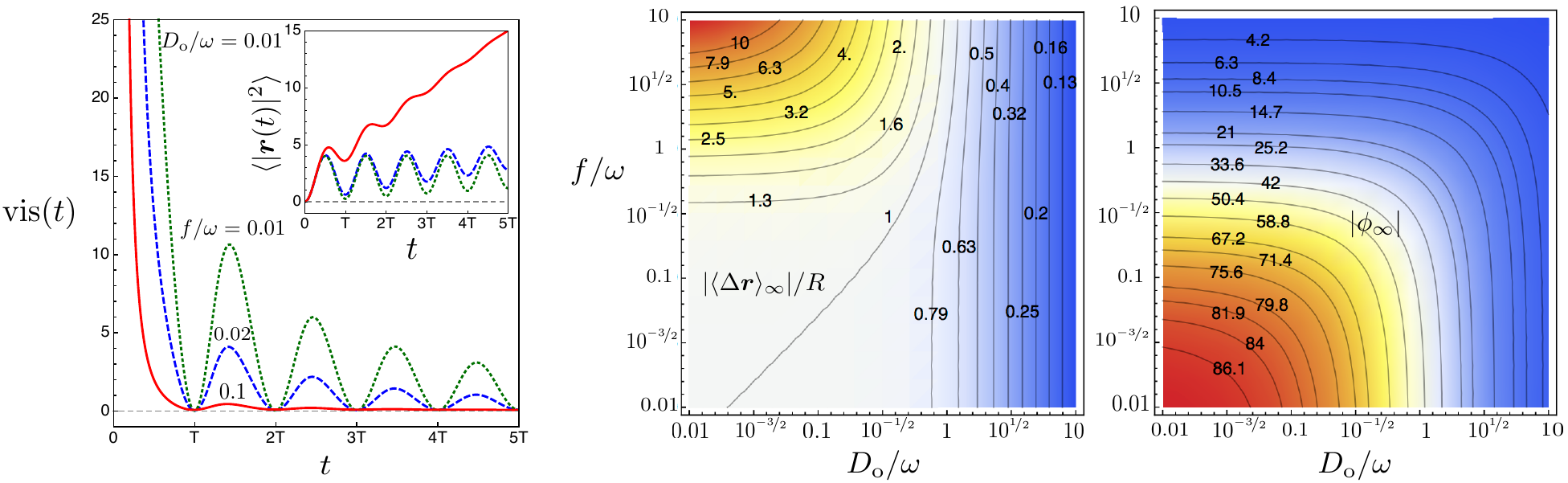}
\end{center}
\vspace{-15pt}
\caption{
(color online)
Left: The visibility ratio 
$\mathrm{vis}(t) = {|\langle{\bm r}(t)\rangle|^2}/{\mathrm{var}({\bm r}(t))}$
measures the ease of discerning the mean behavior of ${\bm r}(t)$ against the
background of its statistical spread.
It drops to near zero more quickly the larger $f$ and $\Dort$.
Inset: mean square displacement reaches the diffusive regime faster as 
$(f+\Dort)/\omega$ increases. 
Middle: 
$|\Drinf| /R =  2 (\Deff/R^2) \sqrt{\omega^{-2}+(\Dort+2f)^{-2}}$ 
as function of $\Dort/\omega$ and $f/\omega$.
Right:
$|\phi_\infty| = \tan^{-1}[\omega/(\Dort+2f)]$, the magnitude of the angle between
$\Drinf$ and $\unitv_0$, as function of $\Dort/\omega$ and 
$f/\omega$.
}
\label{fig:r-inf-phi-inf} 
\end{figure*}
%%%%%%%%%%%

The effective diffusivity $\Deff$ tells us about the asymptotic behavior of the 
mean-squared displacement. At finite times, there are corrections which we will
discuss later. But, more significant for the subject of this Letter 
is the mean displacement vector, given by
\begin{align}
\langle \Delta  {\bm r}(\, {t}\,) \rangle  
& = v \left[\km^{-1} \left({\cal I} - e^{-\km {t}} \right)\repere(0)\right]\cdot \unitv_0
\nonumber \\
&= 
\Drinf - 2\frac{D_{\text{eff}}}{R} {\bm G}(t)
e^{-(\Dort +f) t}, 
\label{eq:displacementvecRT}
\end{align}
where the asymptotic value is
\begin{equation}
\label{eq:asymptotic-mean-displacement}
\Drinf = 2 \frac{D_{\text{eff}} }{R}
\left( \frac{\unitv_0}{\omega} - \frac{\unitp_0}{\Dort+2f} \right).
\end{equation}
The special case of this result for no flipping has been derived previously~\cite{VanTeeffelen}.
Translational diffusion does not affect $\langle \Delta  {\bm r}(\, {t}\,) \rangle$.
Instead, it reflects the interaction of {\em orientational\/} 
diffusion, chirality reversal, and powered rotation.
The second term in the final expression of Eq.~(\ref{eq:displacementvecRT}),
which represents a transient, will be considered momentarily. 
To understand the expression (\ref{eq:asymptotic-mean-displacement}) for the
asymptotic mean displacement, it's helpful to unpack it a little in the 
low-noise limit. Expanding to first order in $\Dort$ and $f$,
we find 
\begin{equation}
\frac{\Drinf}{R} \approx -\unitp_0 + 
{\omega}^{-1}(\Dort+2f)\unitv_0.
\end{equation}
In the limit of vanishing noise, $\Drinf$ tends to the time-average position 
for the deterministic motion on a circle. 
In the presence of noise, there is a deviation, but in 
the direction of the initial velocity.

Turning to the second (transient) term in Eq.~(\ref{eq:displacementvecRT}),
with the abbreviation $\alpha = \sqrt{\omega^2-f^2}$, the vector ${\bm G}(t)$ is
\begin{align}
\label{eq:G-vector}
{\bm G}(t)
&=
\left[\cos \alpha t 
 +
\left(f - \frac{\omega^2}{\Dort+ 2f} \right)
\frac{ \sin \alpha t}{\alpha}\right]\frac{\unitv_0}{\omega}
\nonumber \\[.5 em]
&- \left[\cos \alpha t 
+
\left(\Dort+f \right) \frac{ \sin \alpha t}{\alpha}
\right]\frac{\unitp_0 }{\Dort+2f}.
\end{align}
The asymptote $\Drinf$ has a 
more-or-less uniform qualitative behavior upon varying $\Dort$ and $f$,
but the approach to the asymptote is different (see Fig.~\ref{fig:asympt}).
Specifically, there are two distinct regimes for  $f/\omega$.
If $0 \le f < \omega$, ${\bm G}$ is purely oscillatory.
Thus, the norm of $\langle \Delta  {\bm r}(t)\rangle - \Drinf$
is bounded by a constant multiple of the decaying exponential $\exp[-(\Dort+f)t]$.
Within that bound it oscillates, but the oscillation frequency $\alpha$ 
depends on $f$ and goes to zero as $f$ increases to {$\omega$}. 
This is a little surprising; one might have expected that $\omega$ 
itself was the only possible oscillation frequency.
If $\omega < f$, then ${\bm G}$ grows exponentially with rate 
$(f^2-\omega^2)^{1/2}$. Thus,
$|\langle \Delta  {\bm r}(t)\rangle - \Drinf|
\sim \exp\{-[\Dort + f-(f^2-\omega^2)^{1/2}]t\}$ and the approach is 
non-oscillatory. At fixed $\Dort$, the approach rate has a cusp
at $f=\omega$, and tends to $\Dort$ for both $f=0$ and $f\approx\infty$.

Figure~\ref{fig:spirals} depicts traces of the clockwise rotors' mean trajectory 
conditioned on $\repere(0) = [-\hat{x},\hat{y},-\hat{z}]^T$
 for a range of
values of $\Dort/\omega$ and $f/\omega$, according to the
theory just developed. Although the temporal aspect is lost,
many of the features we have discussed can be seen there.
If $\Dort = f = 0$, then the time average of $\Delta{\bm r}(t)$
is simply $-\unitp_0$. A small amount of noise should cause $\Drinf$ to
deviate by only a small amount from that limit.
More precisely, according to a formula derived earlier, $\Drinf$ should 
move up from the initial orbit center by $R(\Dort+2f)/\omega$; 
Fig.~\ref{fig:spirals2} bears this out up to values of $f \approx \omega$. 
The number of visible oscillations decreases very
rapidly with increasing noise. This is partly due to the increased damping
and partly due to the decreased oscillation frequency [$\alpha$ in Eq.~(\ref{eq:G-vector})].
For $f \gtrsim \omega$, oscillations no longer occur.
Movies contained in Supplementary Information show simulated particle ensembles,
along with their empirical mean positions; thereby, the temporal aspects can be 
better appreciated.

The orderly and revealing behavior of the mean displacement
 is very difficult to
discern in a single trajectory without the sort of special processing we have used.
This difficulty can be quantified, using 
the {\it visibility\/} of the mean displacement $\langle \Delta{\bm r}(t)\rangle$,
defined as
\begin{equation}
\mathrm{vis}(t) = 
\frac{|\langle{\bm r}(t)\rangle|^2}{\mathrm{var}({\bm r}(t))},
\end{equation}
where
$\mathrm{var}({\bm r}(t)) = \langle|{\bm r}(t)|^2\rangle - |\langle{\bm r}(t)\rangle|^2$
is the variance of the position at time $t$. If vis$(t)$ is very small, we cannot expect
to directly discern the mean behavior even in a small ensemble.
The left panel of Fig.~\ref{fig:r-inf-phi-inf} shows plots of visibility at a range 
of 
$f/\omega$
values for $\Dort=0.01\,\omega$. For $\Dort + f \sim 10^{-2}\omega$,
the mean displacement is comparable to the spreading width, and vis$(t)$ takes several
periods to degrade.
At larger values, $\Dort + f \gtrsim 10^{-1}\omega$, the effect is much weaker and 
visibility drops to near zero in less than one period.

Now we turn to a closer look at the long-time asymptote $\Drinf$ of the
mean displacement.
Although, as $f$ increases, $\Drinf$ moves away from the ideal orbit center
in the direction of $\unitv_0$, it does not do so indefinitely; the $f\to\infty$
limit is $\omega R\unitv_0/\Dort$ $[= {\bm v}(0)/\Dort]$.
The center and right panels of Fig.~\ref{fig:r-inf-phi-inf} show details of
the behavior of both the norm 
$|\Drinf|  =  2 (\Deff/R) \sqrt{\omega^{-2}+(\Dort+2f)^{-2}}$
and the absolute value $|\phi_\infty| = \tan^{-1}[\omega/(\Dort+2f)]$
of the angle $\Drinf$ makes with $\unitv_0$ (``chiral angle'').
As the flipping rate $f$ or orientational diffusivity $\Dort$ increases,
$|\phi_\infty|$ decreases. Since $\Drinf$ depends on $f$ and $\Dort$
only through their ratio with $\omega$, this limit can equivalently be
thought of as $\omega\to 0$. From that perspective, the behavior is
understandable as the circular (noise-free) trajectory degenerates
to a straight line.
However, in the experimentally relevant range, $\Dort/\omega \sim 10^{-2}-10^{-1}$,
we always find a non-negligible chiral angle. In the limit of weak noise, $|\Drinf|$
is of the order of the radius of the deterministic trajectory.
At high orientational diffusion, the particle changes its direction
rapidly and therefore $|\Drinf|$ is smaller than 
the radius of the orbit. For $\Dort \ll \omega$ and $f \gg \omega$ the motor
effectively acts more like a rectilinear motor than a rotor, thus the magnitude 
of displacement is much larger than the radius of the orbit.

The asymptotic chiral angle is different. At low noise it is natural to preserve the 
chiral nature; the flipping rate is low and the deviation from circular trajectory 
is small. $|\phi_\infty|$ is very small both when $\Dort \gg \omega$ and 
when $f \gg \omega$, but for different reasons.
Large $\Dort$ implies that trajectories strongly deviate from the
the chiral deterministic rotation; although chirality is preserved, its 
expression is very weak.
With large $f$, on the other hand, the chirality itself is alternating
rapidly. The asymptotic chiral angle reflects only the early memory of 
the initial chirality. After that, the rotor averages out to a linear motion.

In conclusion, we have shown that, in the presence of orientational diffusion
and flipping, the expected position as a function of time of a rotary self-propeller,
relative to its time-zero position and velocity, has significant structure related
to the kinematic parameters. Using ensembles synthesized from single experimental
trajectories, this structure is accessible and can be used to determine the
kinematic parameters. The tell-tale spiral can be clearly seen over much shorter
time scales than those required for the long-time effective diffusion to manifest itself.

\begin{acknowledgments}
The authors are grateful to Prof. Wei Wang for his insightful comments and suggestions.
This work was supported by the National Science Foundation under Grant No. DMR-1420620 through the Penn State Center for Nanoscale Science. J.G.G. acknowledges funding by Northern Arizona University's  College of Engineering, Forestry, $\&$ Natural Sciences (CEFNS).
\end{acknowledgments}

\newpage

\begin{widetext}

\section{tadpoles}
The ``tadpole'' microswimmer particles are fabricated the following way: 
a monolayer of 2$\mu$m diameter SiO$_2$ microspheres (Bangs Laboratories, Inc., 
Fishers, IN) was first deposited onto a clean silicon wafer (Si(100)) using  
Langmuir-Blodgett technique. The substrate was then placed into a physical vapor 
deposition (PVD) system and subsequent thin films of 5 nm titanium (Ti) adhesion 
layer followed by a 10 nm platinum (Pt) catalyst were deposited onto the microbeads 
forming half-coated Janus spheres. The substrate was then tilted by an 
in-vacuum motor to an oblique angle of 85$^\circ$ measuring the angle between the 
surface normal and the incident vapor direction. A thick layer of titania (TiO$_2$) 
was then deposited to a thickness of $\sim$8$\mu$m leading to rod-like formations on 
each microbead. The high-incidence angle deposition is known as Glancing Angle 
Deposition (GLAD). The substrate was then removed from the chamber, and the 
tadpole structures were gently removed from the substrate by bath sonication 
suspending them in 18 M$\Omega$ H$_2$O. The colloidal suspension was mixed with 
varying concentrations (\%v/v) of hydrogen peroxide, H$_2$O$_2$, then pipetted onto 
silicon wafers previously cleaned by oxygen (O$_2$) plasma (Harrick Plasma Ithaca, NY). 
The motion of the tadpoles was observed by brightfield microscopy using a Zeiss 
Axiophot microscope in reflection mode with a 40$\times$ or 60$\times$ dry objective 
coupled to a Mikrotron EoSens GE MC 1364 camera (Unterschleissheim, Germany). 
Videos were recorded at 30 frames per second (fps). Particle tracking was achieved 
with the software $ImageJ$ with translational motion and orientation performed by 
the plugins {\it MTrack2} and {\it OrientationJ}, respectively.

\section{dimer}
First, Janus catalytic beads were made by using a spin coater (Laurell Technologies Corp.) to deposit polystyrene colloids (0.1 \% wt suspension in ethanol of 2 µm diameter beads, Duke Scientific) onto a clean glass slide.  Spin coating conditions were chosen to generate a separate non-touching distribution of colloids (typical conditions: 30 second spin, 2000 rpm, 100 µL dispensed onto spinning substrate).  These glass slides were then subject to directional platinum metal (Agar scientific, 99.9\%) evaporation using a Moorfield Minilab 80 e-beam evaporator (5 nm coating thickness, monitored using a quartz-crystal oscillator).  Damp lens tissue (Whatman) was then used to transfer the metallised colloids from the glass slide into a solution containing hydrogen peroxide (20 \% w/v).  The colloids were incubated for a few days in this solution, during which time agglomerates were observed to form, including the two body swimmers that were investigated in this paper.

In order to explore chiral diffusion phenomena, the suspension of agglomerated swimmers prepared above was diluted to give a 10\% w/v hydrogen peroxide concentration, and then placed into a low volume rectangular glass cuvette (Hellma).  A Nikon Eclipse ME600 microscope operating in transmission mode was used to directly observe the movement of the colloids.  Focus was arranged to ensure that only colloids remaining in close proximity to one of the planar surfaces of the cuvette were investigated.  A camera attached to the microscope (Pixelink PL-742) was used to record videos of the two body agglomerates motion (duration up to 1 hour, frame rate 3-15 Hz).  These videos were subject to automated image analysis using a threshold algorithm to determine the centre of mass for each colloid in each frame with sub-pixel accuracy, output as time-stamped x,y trajectory?s (custom software developed using the National Instruments Labview platform).  Estimates of angular and propulsive velocities for each trajectory were obtained as described in reference 1.  Briefly, MSD versus time plots were generated from the trajectory data, and then the first 1-5 seconds of these were fitted to an analytical expression determine all the relevant motion parameters.  Additionally, the long-axis orientation of the two body swimmer was determined using image analysis, which allowed a second estimated for angular velocity via MSD analysis of the orientation changes as a function of time.

\section{Videos}

\begin{enumerate}

\item vid-fig2.m4v: theoretical trajectories and simulation of an ensemble of nanorotors 
with identical initial position, velocity and chirality. The movie corresponds to 
$D_\text{o}/\omega = 0.1$ and $f/\omega  = 0.0, 0,2, 0.5, 0.7, 1, 1.2$ as in Fig.~1 of the main 
text.

\item vid-NumPtcl.m4v: theoretical trajectories and simulated ensembles and means for
ensemble sizes from 10 to 1000 for $D_\text{o}/\omega = 0.1$ and $f/\omega = 0.01$.

\item vid-trans.m4v: video demonstrating that passive translational diffusivity does not 
affect the shape of the spiral. 

\end{enumerate}

\end{widetext}

\end{document}